%% file: main.tex
\pdfoutput=1
\documentclass[12pt,a4paper]{article}

\usepackage{ifthen} 
\newboolean{pdflatex}
\setboolean{pdflatex}{true} 

\newboolean{articletitles}
\setboolean{articletitles}{true} 

\newboolean{uprightparticles}
\setboolean{uprightparticles}{false} 

\def\paperauthors{LHCb collaboration} 
\def\paperasciititle{Updated search for Bc decays to two charm mesons} 
\def\papertitle{Updated search for \Bc decays to two charm mesons} 
\def\paperkeywords{{High Energy Physics}, {LHCb}} 
\def\papercopyright{\the\year\ CERN for the benefit of the LHCb collaboration} 
\def\paperlicence{CC BY 4.0 licence}
\def\paperlicenceurl{https://creativecommons.org/licenses/by/4.0/}

\input{preamble}
\usepackage{longtable} 
\usepackage{amsmath}
\usepackage{siunitx}
\def\BporBc{{\ensuremath{\B^+_{(\cquark)}}}\xspace}

\def\DporDs{{\ensuremath{\D^+_{(\squark)}}}\xspace}

\def\DporDsorDstar{{\ensuremath{\D^{(*)+}_{(\squark)}}}\xspace}
\def\DmorDsorDstar{{\ensuremath{\D^{(*)-}_{(\squark)}}}\xspace}
\def\DporDstar{{\ensuremath{\D^{(*)+}}}\xspace}
\def\DzorDzb{{\ensuremath{\kern 0.18em\optbar{\kern -0.18em D}{}^0}}\xspace}
\def\DzorDzbstar{{{\ensuremath\kern 0.18em\optbar{\kern -0.18em D}{}^{*0}}}\xspace}
\def\DzorDzborstar{{{\ensuremath\kern 0.18em\optbar{\kern -0.18em D}{}^{(*)0}}}\xspace}

\def\inlfcfu{{\ensuremath{f_c/f_u}}\xspace}
\def\inlfcfufd{{\ensuremath{f_c/(f_u+f_d)}}\xspace}
\def\DKpi{{\ensuremath{\mbox{\decay{\Dz}{K^-\pi^+}}}}\xspace}

\def\DKpipipi{{\ensuremath{\mbox{\decay{\Dz}{K^-\pi^+\pi^-\pi^+}}}}\xspace}

\def\DzorDzstar{{\ensuremath{\D^{(*)0}}}\xspace}

\def\DporDsstar{{\ensuremath{\D^{*+}_{(\squark)}}}\xspace}
\def\DporDsorstar{{\ensuremath{\D^{(*)+}_{(\squark)}}}\xspace}

\def\Dpstar{{\ensuremath{\D^{*+}}}\xspace}

\def\K3pi{{\ensuremath{K3\pi}}\xspace}
\def\pizgamma{{\ensuremath{X^0}}\xspace}

\begin{document}

\renewcommand{\thefootnote}{\fnsymbol{footnote}}
\setcounter{footnote}{1}

\input{title-LHCb-PAPER}


\renewcommand{\thefootnote}{\fnsymbol{footnote}}
\setcounter{footnote}{2}

\cleardoublepage


\pagestyle{plain} 
\setcounter{page}{1}
\pagenumbering{arabic}


\input{body}

\input{acknowledgements}



\newpage

\addcontentsline{toc}{section}{References}
\bibliographystyle{LHCb}
\bibliography{main,standard,LHCb-PAPER,LHCb-CONF,LHCb-DP,LHCb-TDR}

\newpage
\input{Authorship_LHCb-PAPER-2021-023}



\end{document}

%% file: preamble.tex
\usepackage[top=1in, bottom=1.25in, left=1in, right=1in]{geometry}

\usepackage{microtype}
\usepackage{lineno}  
\usepackage{xspace} 
\usepackage{caption} 

\usepackage{graphicx}  
\usepackage{color}
\usepackage{colortbl}
\graphicspath{{./figs/}} 

\usepackage{amsmath} 
\usepackage{amssymb}
\usepackage{amsfonts}
\usepackage{upgreek} 

\newcommand*\patchAmsMathEnvironmentForLineno[1]{%
\expandafter\let\csname old#1\expandafter\endcsname\csname #1\endcsname
\expandafter\let\csname oldend#1\expandafter\endcsname\csname
end#1\endcsname
 \renewenvironment{#1}%
   {\linenomath\csname old#1\endcsname}%
   {\csname oldend#1\endcsname\endlinenomath}%
}
\newcommand*\patchBothAmsMathEnvironmentsForLineno[1]{%
  \patchAmsMathEnvironmentForLineno{#1}%
  \patchAmsMathEnvironmentForLineno{#1*}%
}
\AtBeginDocument{%
\patchBothAmsMathEnvironmentsForLineno{equation}%
\patchBothAmsMathEnvironmentsForLineno{align}%
\patchBothAmsMathEnvironmentsForLineno{flalign}%
\patchBothAmsMathEnvironmentsForLineno{alignat}%
\patchBothAmsMathEnvironmentsForLineno{gather}%
\patchBothAmsMathEnvironmentsForLineno{multline}%
\patchBothAmsMathEnvironmentsForLineno{eqnarray}%
}

\usepackage{hyperxmp}

\usepackage[pdftex,
            pdfauthor={\paperauthors},
            pdftitle={\paperasciititle},
            pdfkeywords={\paperkeywords},
            pdfcopyright={Copyright (C) \papercopyright},
            pdflicenseurl={\paperlicenceurl}]{hyperref}

\usepackage[colorinlistoftodos,textsize=scriptsize]{todonotes}

\usepackage[bottom,flushmargin,hang,multiple]{footmisc}

\usepackage[all]{hypcap} 

\input{lhcb-symbols-def} 

\usepackage{cite} 
\usepackage{mciteplus}

%% file: lhcb-symbols-def.tex
\usepackage{xspace} 
\usepackage{upgreek}

\def\lhcb   {\mbox{LHCb}\xspace}

\def\MagUp {\mbox{\em Mag\kern -0.05em Up}\xspace}

\ifthenelse{\boolean{uprightparticles}}%
{

 \def\Pmu         {\ensuremath{\upmu}\xspace}                 
 \def\Pnu         {\ensuremath{\upnu}\xspace}                 
                  
 \def\Ppi         {\ensuremath{\uppi}\xspace}

 \def\Ppsi        {\ensuremath{\uppsi}\xspace}

 \def\PDelta      {\ensuremath{\Delta}\xspace}                 
 \def\PXi         {\ensuremath{\Xi}\xspace}                 
 \def\PLambda     {\ensuremath{\Lambda}\xspace}                 
 \def\PSigma      {\ensuremath{\Sigma}\xspace}                 
 \def\POmega      {\ensuremath{\Omega}\xspace}                 
 \def\PUpsilon    {\ensuremath{\Upsilon}\xspace}

 \def\PB      {\ensuremath{\mathrm{B}}\xspace}                 
                  
 \def\PD      {\ensuremath{\mathrm{D}}\xspace}

 \def\PJ      {\ensuremath{\mathrm{J}}\xspace}                 
 \def\PK      {\ensuremath{\mathrm{K}}\xspace}

 \def\Pb      {\ensuremath{\mathrm{b}}\xspace}                 
 \def\Pc      {\ensuremath{\mathrm{c}}\xspace}                 
 \def\Pd      {\ensuremath{\mathrm{d}}\xspace}

 \def\Pi      {\ensuremath{\mathrm{i}}\xspace}

 \def\Ps      {\ensuremath{\mathrm{s}}\xspace}                 
                  
 \def\Pu      {\ensuremath{\mathrm{u}}\xspace}

 \def\thebaroffset{0.0em}
}
{

 \def\Pmu         {\ensuremath{\mu}\xspace}                 
 \def\Pnu         {\ensuremath{\nu}\xspace}                 
                  
 \def\Ppi         {\ensuremath{\pi}\xspace}

 \def\Ppsi        {\ensuremath{\psi}\xspace}                 
                  
 \mathchardef\PDelta="7101
 \mathchardef\PXi="7104
 \mathchardef\PLambda="7103
 \mathchardef\PSigma="7106
 \mathchardef\POmega="710A
 \mathchardef\PUpsilon="7107
                  
 \def\PB      {\ensuremath{B}\xspace}                 
                  
 \def\PD      {\ensuremath{D}\xspace}

 \def\PJ      {\ensuremath{J}\xspace}                 
 \def\PK      {\ensuremath{K}\xspace}

 \def\Pb      {\ensuremath{b}\xspace}                 
 \def\Pc      {\ensuremath{c}\xspace}                 
 \def\Pd      {\ensuremath{d}\xspace}

 \def\Pi      {\ensuremath{i}\xspace}

 \def\Ps      {\ensuremath{s}\xspace}                 
                  
 \def\Pu      {\ensuremath{u}\xspace}

 \def\thebaroffset{0.18em}
}
\newcommand{\offsetoverline}[2][\thebaroffset]{\kern #1\overline{\kern -#1 #2}}%

\makeatletter
\ifcase \@ptsize \relax
  \newcommand{\miniscule}{\@setfontsize\miniscule{4}{5}}
\or
  \newcommand{\miniscule}{\@setfontsize\miniscule{5}{6}}
\or
  \newcommand{\miniscule}{\@setfontsize\miniscule{5}{6}}
\fi
\makeatother

\DeclareRobustCommand{\optbar}[1]{\shortstack{{\miniscule (\rule[.5ex]{1.25em}{.18mm})}
  \\ [-.7ex] $#1$}}

\def\mup        {{\ensuremath{\Pmu^+}}\xspace}

\def\neu        {{\ensuremath{\Pnu}}\xspace}

\def\neum       {{\ensuremath{\neu_\mu}}\xspace}

\def\uquark    {{\ensuremath{\Pu}}\xspace}

\def\dquark    {{\ensuremath{\Pd}}\xspace}

\def\squark    {{\ensuremath{\Ps}}\xspace}

\def\cquark    {{\ensuremath{\Pc}}\xspace}

\def\bquark    {{\ensuremath{\Pb}}\xspace}
\def\bquarkbar {{\ensuremath{\overline \bquark}}\xspace}

\def\pion   {{\ensuremath{\Ppi}}\xspace}
\def\piz    {{\ensuremath{\pion^0}}\xspace}
\def\pip    {{\ensuremath{\pion^+}}\xspace}
\def\pim    {{\ensuremath{\pion^-}}\xspace}

\def\kaon    {{\ensuremath{\PK}}\xspace}

\def\KorKbar {\kern \thebaroffset\optbar{\kern -\thebaroffset \PK}{}\xspace}

\def\Kp      {{\ensuremath{\kaon^+}}\xspace}
\def\Km      {{\ensuremath{\kaon^-}}\xspace}

\def\Dbar    {{\ensuremath{\offsetoverline{\PD}}}\xspace}
\def\D       {{\ensuremath{\PD}}\xspace}

\def\DorDbar {\kern \thebaroffset\optbar{\kern -\thebaroffset \PD}\xspace}
\def\Dz      {{\ensuremath{\D^0}}\xspace}
\def\Dzb     {{\ensuremath{\Dbar{}^0}}\xspace}
\def\Dp      {{\ensuremath{\D^+}}\xspace}
\def\Dm      {{\ensuremath{\D^-}}\xspace}

\def\DpDm    {\ensuremath{\Dp {\kern -0.16em \Dm}}\xspace}

\def\Dstarz  {{\ensuremath{\D^{*0}}}\xspace}
\def\Dstarzb {{\ensuremath{\Dbar{}^{*0}}}\xspace}

\def\Dstarp  {{\ensuremath{\D^{*+}}}\xspace}

\def\Ds      {{\ensuremath{\D^+_\squark}}\xspace}

\def\Dss     {{\ensuremath{\D^{*+}_\squark}}\xspace}

\def\B       {{\ensuremath{\PB}}\xspace}

\def\BorBbar {\kern \thebaroffset\optbar{\kern -\thebaroffset \PB}\xspace}
\def\Bz      {{\ensuremath{\B^0}}\xspace}

\def\Bd      {{\ensuremath{\B^0}}\xspace}

\def\BdorBdbar {\kern \thebaroffset\optbar{\kern -\thebaroffset \Bd}\xspace}
\def\Bu      {{\ensuremath{\B^+}}\xspace}

\def\Bp      {{\ensuremath{\Bu}}\xspace}

\def\Bs      {{\ensuremath{\B^0_\squark}}\xspace}

\def\BsorBsbar {\kern \thebaroffset\optbar{\kern -\thebaroffset \Bs}\xspace}
\def\Bc      {{\ensuremath{\B_\cquark^+}}\xspace}

\def\jpsi     {{\ensuremath{{\PJ\mskip -3mu/\mskip -2mu\Ppsi}}}\xspace}

\def\Y#1S{\ensuremath{\PUpsilon{(#1S)}}\xspace}

\def\LorLbar     {\kern \thebaroffset\optbar{\kern -\thebaroffset \PLambda}\xspace}

\def\BF         {{\ensuremath{\mathcal{B}}}\xspace}

\newcommand{\decay}[2]{\ensuremath{#1\!\to #2}\xspace} 

\def\to                 {\ensuremath{\rightarrow}\xspace}

\def\order   {{\ensuremath{\mathcal{O}}}\xspace}

\def\CP                {{\ensuremath{C\!P}}\xspace}

\def\Vud  {{\ensuremath{V_{\uquark\dquark}^{\phantom{\ast}}}}\xspace}
\def\Vcd  {{\ensuremath{V_{\cquark\dquark}^{\phantom{\ast}}}}\xspace}

\def\Vubs  {{\ensuremath{V_{\uquark\bquark}^\ast}}\xspace}
\def\Vcbs  {{\ensuremath{V_{\cquark\bquark}^\ast}}\xspace}

\def\AT#1     {\ensuremath{A_{\mathrm{T}}^{#1}}\xspace}           

\def\C#1      {\ensuremath{\mathcal{C}_{#1}}\xspace}                       
\def\Cp#1     {\ensuremath{\mathcal{C}_{#1}^{'}}\xspace}                    
\def\Ceff#1   {\ensuremath{\mathcal{C}_{#1}^{\mathrm{(eff)}}}\xspace}        
\def\Cpeff#1  {\ensuremath{\mathcal{C}_{#1}^{'\mathrm{(eff)}}}\xspace}       
\def\Ope#1    {\ensuremath{\mathcal{O}_{#1}}\xspace}                       
\def\Opep#1   {\ensuremath{\mathcal{O}_{#1}^{'}}\xspace}                    

\newcommand{\nospaceunit}[1]{\ensuremath{\text{#1}}}       
\newcommand{\aunit}[1]{\ensuremath{\text{\,#1}}}       

\newcommand{\tev}{\aunit{Te\kern -0.1em V}\xspace}
\newcommand{\gev}{\aunit{Ge\kern -0.1em V}\xspace}
\newcommand{\mev}{\aunit{Me\kern -0.1em V}\xspace}
\newcommand{\kev}{\aunit{ke\kern -0.1em V}\xspace}
\newcommand{\ev}{\aunit{e\kern -0.1em V}\xspace}
 
\newcommand{\mevc}{\ensuremath{\aunit{Me\kern -0.1em V\!/}c}\xspace}
\newcommand{\gevc}{\ensuremath{\aunit{Ge\kern -0.1em V\!/}c}\xspace}
\newcommand{\mevcc}{\ensuremath{\aunit{Me\kern -0.1em V\!/}c^2}\xspace}
\newcommand{\gevcc}{\ensuremath{\aunit{Ge\kern -0.1em V\!/}c^2}\xspace}

\def\mum  {\ensuremath{\,\upmu\nospaceunit{m}}\xspace}

\def\fb   {\ensuremath{\aunit{fb}}\xspace}
\def\invfb   {\ensuremath{\fb^{-1}}\xspace}

\def\order{{\ensuremath{\mathcal{O}}}\xspace}
\newcommand{\chisq}{\ensuremath{\chi^2}\xspace}

\newcommand{\chisqip}{\ensuremath{\chi^2_{\text{IP}}}\xspace}

\def\gsim{{~\raise.15em\hbox{$>$}\kern-.85em
          \lower.35em\hbox{$\sim$}~}\xspace}
\def\lsim{{~\raise.15em\hbox{$<$}\kern-.85em
          \lower.35em\hbox{$\sim$}~}\xspace}

\def\pt         {\ensuremath{p_{\mathrm{T}}}\xspace}

\def\ptot       {\ensuremath{p}\xspace}

\def\dllkpi     {\ensuremath{\mathrm{DLL}_{\kaon\pion}}\xspace}

\def\mrad{\aunit{mrad}\xspace}

\def\bcvegpy    {\mbox{\textsc{Bcvegpy}}\xspace}

\def\evtgen     {\mbox{\textsc{EvtGen}}\xspace}

\def\geant      {\mbox{\textsc{Geant4}}\xspace}

\def\photos     {\mbox{\textsc{Photos}}\xspace}

\def\pythia     {\mbox{\textsc{Pythia}}\xspace}

\def\tell1  {TELL1\xspace}
\def\ukl1   {UKL1\xspace}



%% file: title-LHCb-PAPER.tex

\begin{titlepage}
\pagenumbering{roman}

\vspace*{-1.5cm}
\centerline{\large EUROPEAN ORGANIZATION FOR NUCLEAR RESEARCH (CERN)}
\vspace*{1.5cm}
\noindent
\begin{tabular*}{\linewidth}{lc@{\extracolsep{\fill}}r@{\extracolsep{0pt}}}
\ifthenelse{\boolean{pdflatex}}
{\vspace*{-1.5cm}\mbox{\!\!\!\includegraphics[width=.14\textwidth]{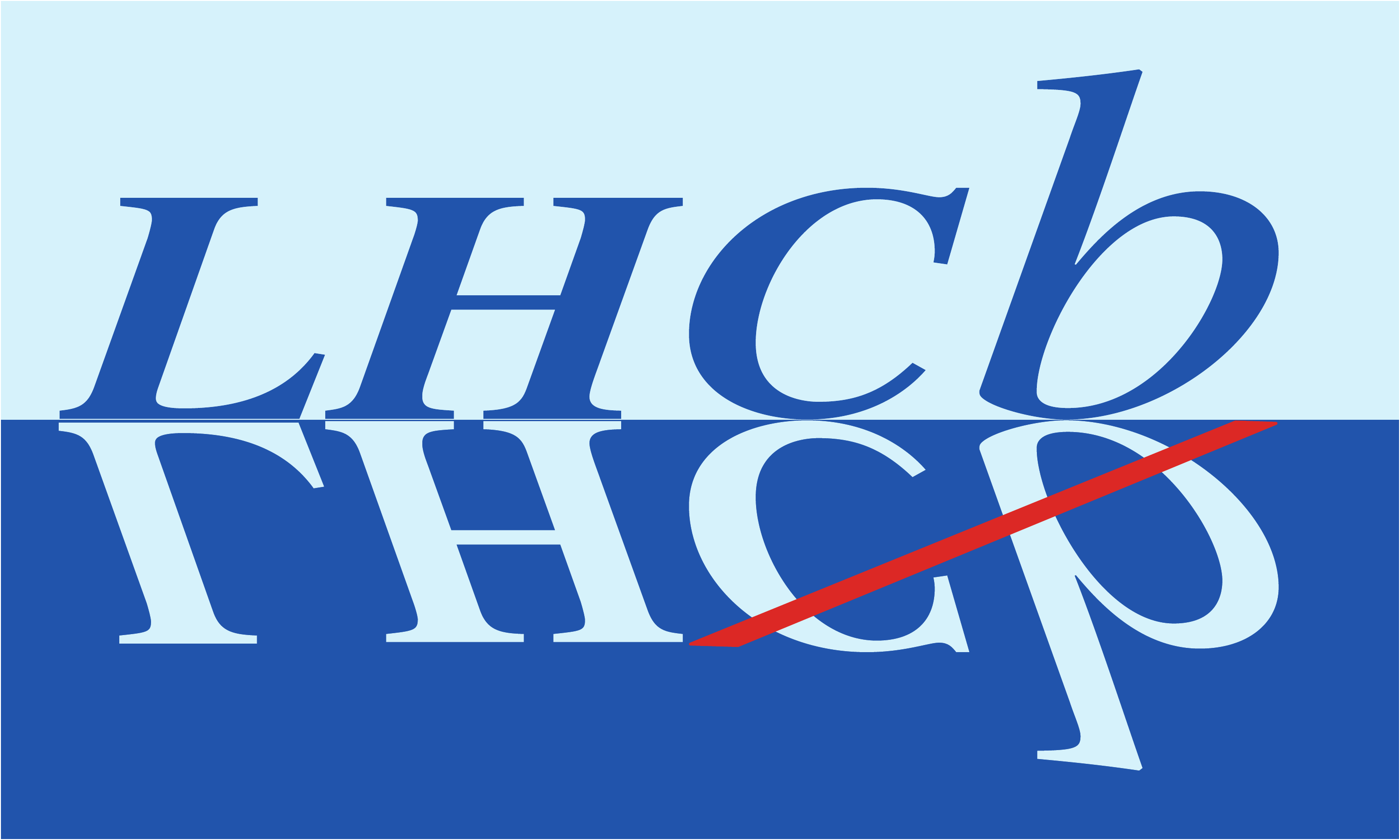}} & &}%
{\vspace*{-1.2cm}\mbox{\!\!\!\includegraphics[width=.12\textwidth]{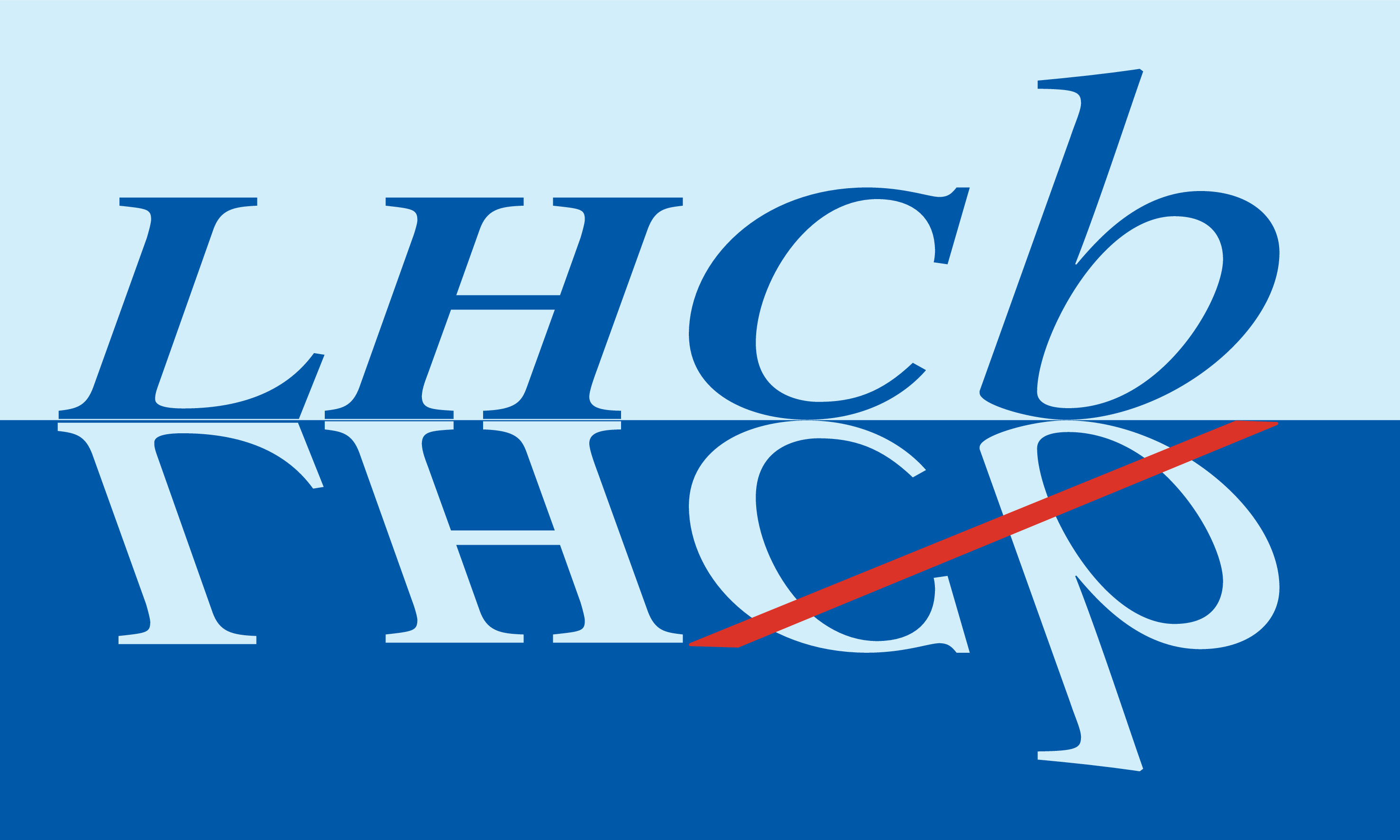}} & &}%
\\
 & & CERN-EP-2021-154 \\  
 & & LHCb-PAPER-2021-023 \\  
 & & December 17, 2021 \\ 
\end{tabular*}

\vspace*{4.0cm}

{\normalfont\bfseries\boldmath\huge
\begin{center}
  \papertitle 
\end{center}
}

\vspace*{2.0cm}

\begin{center}
\paperauthors\footnote{Authors are listed at the end of this paper.}
\end{center}

\vspace{\fill}

\begin{abstract}
  \noindent
A data set corresponding to an integrated luminosity of $9 \invfb$ of proton-proton collisions collected by the LHCb experiment has been analysed to search for \mbox{\decay{\Bc}{\DporDsorstar\DzorDzborstar}} decays.
The decays are fully or partially reconstructed,
where one or two missing neutral pions or photons from the decay of an excited charm meson are allowed.
Upper limits for the branching fractions, normalised to \Bp decays 
to final states with similar topologies, are obtained for sixteen \Bc decay modes.
For the decay \decay{\Bc}{\Ds\Dzb}, an excess with a significance of 3.4 standard deviations is found.
  
\end{abstract}

\vspace*{2.0cm}

\begin{center}
  Published in JHEP  12 (2021) 117  
\end{center}

\vspace{\fill}

{\footnotesize 
\centerline{\copyright~\papercopyright. \href{\paperlicenceurl}{\paperlicence}.}}
\vspace*{2mm}

\end{titlepage}


\newpage
\setcounter{page}{2}
\mbox{~}
%
%
%
%

%% file: body.tex
\section{Introduction}
\label{sec:Introduction}

Heavy-flavour states with $b$ quarks 
are characterised by a relatively long lifetime and
a large number of decay channels,
and allow for 
highly sensitive studies
of charge and parity (\CP) symmetry violation and quantum-loop induced amplitudes.
In the \Bc meson, a \bquarkbar quark is accompanied by a charm quark, \cquark,
forming a system where decays of both the beauty and the charm 
quark, as well as weak annihilation processes, contribute to the decay amplitude~\cite{LHCb-PAPER-2016-058}.

\begin{figure}[b]
    \centering
        \includegraphics[trim=0.0cm 0.0cm 0.0cm 2.0cm, width=0.48\textwidth]{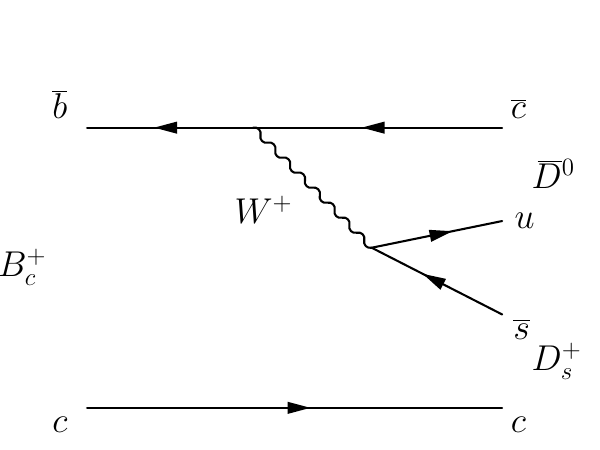}
        \includegraphics[trim=0.0cm 0.0cm 0.0cm 2.0cm, width=0.48\textwidth]{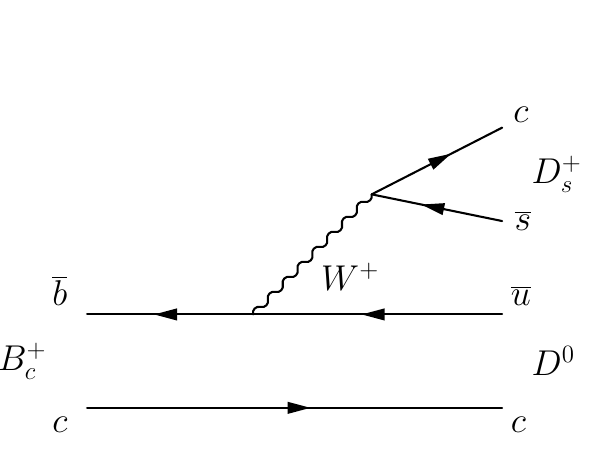}
    \caption{Diagrams for the (left) CKM-favoured, colour-suppressed \mbox{\decay{\Bc}{\Ds\Dzb}}
    and (right) CKM-suppressed, colour-favoured \mbox{\decay{\Bc}{\Ds\Dz}} decays.
 }
    \label{fig:Bc2DsDz}
\end{figure}

Transition amplitudes between up-type quarks and down-type quarks are described by the Cabibbo-Kobayashi-Maskawa (CKM) 
quark-mixing matrix~\cite{Cabibbo:1963yz,Kobayashi:1973fv}.
Figure~\ref{fig:Bc2DsDz} illustrates the CKM-favoured, but colour-suppressed \mbox{\decay{\Bc}{\Ds\Dzb}} decay (unless specified otherwise, charge conjugation is implied throughout this article) and the CKM-suppressed, but colour-favoured \mbox{\decay{\Bc}{\Ds\Dz}} decay, which are expected to have similar amplitudes.
This may result in a large, $\order(1)$, \CP asymmetry for final states that are common between \Dz and \Dzb decays.
Consequently decays of \Bc mesons to two charm mesons, \mbox{\decay{\Bc}{\DporDs\DzorDzb}},
have been proposed to measure the angle $\gamma\equiv {\rm arg}(-\Vud\Vubs/\Vcd\Vcbs)$~\cite{Masetti:1992in,Fleischer:2000pp,Giri:2001be,Giri:2006cw}, one of the key parameters of the CKM matrix.
Presently, the most precise determinations of $\gamma$ come from measurements of
the \CP asymmetry in \mbox{\decay{\Bp}{\DzorDzb\Kp}} decays~\cite{LHCb-PAPER-2020-036,LHCb-PAPER-2020-019}.

Predicted branching fractions of 
\Bc decays to two charm mesons
~\cite{AbdEl-Hady:1999jux,Rui:2012qq,Kiselev:2003ds,Ivanov:2002un,Ivanov:2006ni}
are listed in Table~\ref{table:bfestimate}. 
Final-state interactions may result in an enhancement of \mbox{\decay{\Bc}{\Dp\DzorDzb}} decay rates~\cite{Mohammadi:2021snr}.
Moreover, contributions from physics beyond the Standard Model could potentially affect fully hadronic \B
decays~\cite{Bordone:2020gao,Iguro:2020ndk,Cai:2021mlt}.

\begin{table}[tb]
  \caption{
    Predicted branching fractions of \Bc decays to two charm mesons, in units of $10^{-6}$.}
  \begin{center}\begin{tabular}{lr@{\:$\pm$\:}lccc}
  \hline
    Channel                & \multicolumn{2}{c}{Ref.~\cite{Rui:2012qq}} &\multicolumn{1}{c}{Ref.\cite{Kiselev:2003ds}}&\multicolumn{1}{c}{Ref.\cite{Ivanov:2002un}}&\multicolumn{1}{c}{Ref.\cite{Ivanov:2006ni}} \\
      \hline
      \vspace{-0.45cm} \\
    \decay{\Bc}{\Ds\Dzb}     & 2.3 &0.5  & 4.8 & 1.7 & 2.1\\
    \decay{\Bc}{\Ds\Dz}      & 3.0 &0.5  & 6.6 & 2.5 & 7.4\\
    \decay{\Bc}{\Dp\Dzb}     & 32  &7    & 53  & 32  &  33\\
    \decay{\Bc}{\Dp\Dz}      & 0.10&0.02 & 0.32& 0.11& 0.31\\  
    \decay{\Bc}{\Dpstar\Dzb} & 12 &3     & 49  & 17  & 9\\  
    \decay{\Bc}{\Dpstar\Dz}  & 0.09&0.02 & 0.40& 0.38& 0.44\\  
  \hline
  \end{tabular}\end{center}
\label{table:bfestimate}
\end{table}

This article describes a search for sixteen \mbox{\decay{\Bc}{\DporDsorstar\DzorDzborstar}} decay channels,
using proton-proton $(pp)$ collision data collected by the LHCb experiment, corresponding 
to an integrated luminosity of $9\invfb$, 
of which $1\invfb$ was recorded at a centre-of-mass energy 
$\sqrt{s}=7$\tev, $2\invfb$ at $\sqrt{s}=8$\tev and $6\invfb$ at $\sqrt{s}=13$\tev.
The data taken at 7 and $8\tev$ are referred to as Run~1, and the data taken at $13\tev$ as Run~2.
The Run 1 data has previously been analysed and no evidence of \mbox{\decay{\Bc}{\DporDsorstar\DzorDzborstar}} decays was found~\cite{LHCb-PAPER-2017-045}.

Charm mesons are reconstructed in the  
\mbox{\decay{\Dz}{\Km\pip}},
\mbox{\decay{\Dz}{\Km\pip\pim\pip}},
\mbox{\decay{\Dp}{\Km\pip\pip}},
\mbox{\decay{\Ds}{\Kp\Km\pip}}, and
\mbox{\decay{\Dstarp}{\Dz\pip}}
decay modes.
In the decay \mbox{\decay{\Bc}{\Dstarp\DzorDzb}}, 
at least one of the neutral charm mesons
is required to decay as \mbox{\decay{\Dz}{\Km\pip}}.
Partially reconstructed \Bc decays, which involve one or two excited charm mesons producing a photon or a neutral pion in their decay, are also included in the search.
These decays manifest themselves as relatively narrow structures in the mass distributions of the reconstructed final states below the \Bc mass.

The branching fractions, \BF, of \Bc decays to fully reconstructed final states are measured
relative to high-yield \mbox{\decay{\Bp}{\DporDsorDstar\Dzb}} normalisation modes,
\begin{equation}
R(\DporDsorDstar\DzorDzb)\equiv\frac{\raisebox{0.3em}{$f_c$}}{f_u}\frac{\BF(\decay{\Bc}{\DporDsorDstar\DzorDzb})}{\BF(\decay{\Bp}{\DporDsorDstar\Dzb})}
=\frac{N(\decay{\Bc}{\DporDsorDstar\DzorDzb})}{\varepsilon(\decay{\Bc}{\DporDsorDstar\DzorDzb})}
\frac{\varepsilon(\decay{\Bp}{\DporDsorDstar\Dzb})}{N(\decay{\Bp}{\DporDsorDstar\Dzb})},
\label{eq:bfcalc}
\end{equation}
where \inlfcfu is the ratio of the \Bc to \Bp fragmentation fraction,
$N$ denotes the measured \BporBc yields, and $\varepsilon$ represents the detection efficiencies.
The value of \mbox{$\inlfcfufd\cdot \BF(\decay{\Bc}{\jpsi\mup\neum})$} has been measured at centre-of-mass energies of 7 and $13\tev$~\cite{LHCb-PAPER-2019-033}.
Under the assumption of equal production from hadronisation of \Bp and \Bz, $f_u=f_d$,
the value of $\inlfcfu$ is found to be 0.73\% at $\sqrt{s}=7\tev$ and 0.76\% at $\sqrt{s}=13\tev$ with relative uncertainties
of approximately 25\%, dominated by the uncertainty on
the predicted value of \mbox{$\BF(\decay{\Bc}{\jpsi\mup\neum})$},
for which no measurements are available.
 Earlier measurements of \inlfcfu at 7 and $8\tev$ using fully reconstructed \Bc decays found compatible values~\cite{LHCb-PAPER-2012-028,LHCb-PAPER-2014-050}.

The invariant-mass distributions of partially reconstructed \mbox{\decay{\Bc}{\DporDsstar\DzorDzb}} 
and \mbox{\decay{\Bc}{\DporDs\DzorDzbstar}} decays overlap.
Their branching fractions are measured separately by treating the contribution as arising entirely from each decay:

\begin{equation}
    \begin{split}
 R_+'(\DporDs\DzorDzb)&\equiv\frac{\raisebox{0.3em}{$f_c$}}{f_u}\frac{  \BF(\decay{\Bc}{\DporDsstar\DzorDzb}) }{\BF(\decay{\Bp}{\DporDs\Dzb})} \\
&=\frac{N(\decay{\Bc}{\DporDsstar\DzorDzb})}{\varepsilon(\decay{\Bc}{\DporDsstar\DzorDzb})\BF(\decay{\DporDsstar}{\DporDs\pizgamma})}
  \frac{\varepsilon(\decay{\Bp}{\DporDs\Dzb})}{N(\decay{\Bp}{\DporDs\Dzb})},
  \label{eq:bfcalc_1SPDpDs}
    \end{split}
\end{equation}

\begin{equation}
 R_0'(\DporDs\DzorDzb)\equiv\frac{\raisebox{0.3em}{$f_c$}}{f_u}\frac{  \BF(\decay{\Bc}{\DporDs\DzorDzbstar}) }{\BF(\decay{\Bp}{\DporDs\Dzb})}
=\frac{N(\decay{\Bc}{\DporDs\DzorDzbstar})}{\varepsilon(\decay{\Bc}{\DporDs\DzorDzbstar})}
  \frac{\varepsilon(\decay{\Bp}{\DporDs\Dzb})}{N(\decay{\Bp}{\DporDs\Dzb})},
  \label{eq:bfcalc_1SZDpDz}
\end{equation}

 where \pizgamma represents a neutral pion or a photon.
 Decays of \mbox{\decay{\Bc}{\Dpstar\DzorDzbstar}} with 
a fully reconstructed \decay{\Dstarp}{\Dz\pip} decay, and one missing neutral pion or photon from the $\DzorDzbstar$ meson decay, results in measurements of 
\begin{multline}
R'(\Dstarp\DzorDzb)\equiv\frac{\raisebox{0.13em}{$f_c$}}{f_u}\frac{  \BF(\decay{\Bc}{\Dstarp\DzorDzbstar}) }{\BF(\decay{\Bp}{\Dstarp\Dzb})}
= \frac{N(\decay{\Bc}{\Dpstar\DzorDzbstar})}{\varepsilon(\decay{\Bc}{\Dstarp\DzorDzbstar})} \frac{\varepsilon(\decay{\Bp}{\Dstarp\Dzb})}{N(\decay{\Bp}{\Dstarp\Dzb})}.
\label{eq:bfcalc_1S_Dstar}
\end{multline}

The \mbox{\decay{\Bc}{\Dss\DzorDzbstar}} and \mbox{\decay{\Bc}{\Dstarp\DzorDzbstar}} decays can also be observed when both excited charm mesons
decay with either a photon or a neutral pion and neither of the two neutral particles are reconstructed. In such cases, the ratio $R''$ is measured:
\begin{equation}
\begin{split}
R''(\DporDs\DzorDzb)&\equiv\frac{\raisebox{0.3em}{$f_c$}}{f_u}\frac{\BF(\decay{\Bc}{\DporDsstar\DzorDzbstar})}{\BF(\decay{\Bp}{\DporDs\Dzb})}\\
&=\frac{N(\decay{\Bc}{\DporDsstar\DzorDzbstar})}{\varepsilon(\decay{\Bc}{\DporDsstar\DzorDzbstar})\BF(\decay{\DporDsstar}{\DporDs\pizgamma})}
\frac{\varepsilon(\decay{\Bp}{\DporDs\Dzb})}{N(\decay{\Bp}{\DporDs\Dzb})}.
\label{eq:bfcalc_2S}
\end{split}
\end{equation}

In total twenty ratios are measured, corresponding to sixteen \Bc branching fractions,
since \Bc decays with a \Dpstar in the final state are searched for both in fully reconstructed \decay{\Dpstar}{\Dz\pip}
and in partially reconstructed \decay{\Dpstar}{\Dp\pizgamma} decays.

\section{Detector and simulation}
\label{sec:Detector}

The \lhcb detector~\cite{LHCb-DP-2008-001,LHCb-DP-2014-002} is a single-arm forward
spectrometer covering the \mbox{pseudorapidity} range $2<\eta <5$,
designed for the study of particles containing \bquark or \cquark
quarks. The detector includes a high-precision tracking system
consisting of a silicon-strip vertex detector surrounding the $pp$
interaction region~\cite{LHCb-DP-2014-001}, a large-area silicon-strip detector located
upstream of a dipole magnet with a bending power of about
$4{\mathrm{\,Tm}}$, and three stations of silicon-strip detectors and straw
drift tubes~\cite{LHCb-DP-2013-003,LHCb-DP-2017-001}
placed downstream of the magnet.
The tracking system provides a measurement of the momentum, \ptot, of charged particles with
a relative uncertainty that varies from 0.5\% at low momentum to 1.0\% at $200\gevc$.
The minimum distance of a track to a primary $pp$ collision vertex (PV), the impact parameter, 
is measured with a resolution of $(15+29/\pt)\mum$,
where \pt is the component of the momentum transverse to the beam, in\,\gevc.
Different types of charged hadrons are distinguished using information
from two ring-imaging Cherenkov detectors~\cite{LHCb-DP-2012-003}. 
Photons, electrons and hadrons are identified by a calorimeter system consisting of
scintillating-pad and preshower detectors, an electromagnetic
and a hadronic calorimeter. Muons are identified by a
system composed of alternating layers of iron and multiwire
proportional chambers~\cite{LHCb-DP-2012-002}.
The online event selection is performed by a trigger~\cite{LHCb-DP-2012-004}, 
which consists of a hardware stage, based on information from the calorimeter and muon
systems, followed by a software stage, which applies a full event
reconstruction.

At the hardware trigger stage, events are required to have a muon with high \pt or a
  hadron, photon or electron with high transverse energy in the calorimeters. For hadrons,
  the transverse energy threshold is $3.5\gev$.
  The software trigger requires a two-, three- or four-track
  secondary vertex with a significant displacement from any PV. 
  At least one track should have $\pt>1.7\gevc$ and \chisqip with respect to any
  PV greater than 16, where \chisqip is defined as the
  difference in the vertex-fit \chisq of a given PV reconstructed with and
  without the considered particle.
  A multivariate algorithm~\cite{BBDT,LHCb-PROC-2015-018} is used for
  the identification of secondary vertices consistent with the decay
  of a \bquark hadron.

Simulation is used to model the effects of the detector acceptance and the
  imposed selection requirements, as well as for the training of the multivariate selection of the \Bc signals,
 and for establishing the shape of the mass distributions of the signals.
  The \pythia~\cite{Sjostrand:2007gs,*Sjostrand:2006za} package,
  with a specific \lhcb configuration~\cite{LHCb-PROC-2010-056},
  is used to simulate $pp$ collisions with \Bp production.
  For \Bc production, the \bcvegpy~\cite{Chang:2003cq} generator is used,
  interfaced with the Pythia parton shower and hadronisation model.
  Decays of unstable particles
  are described by \evtgen~\cite{Lange:2001uf}, in which final-state
  radiation is generated using \photos~\cite{davidson2015photos}.
  The interaction of the generated particles with the detector, and its response,
  are implemented using the \geant
  toolkit~\cite{Allison:2006ve, *Agostinelli:2002hh} as described in
  Ref.~\cite{LHCb-PROC-2011-006}.
The simulated \Bp production is corrected to match the observed spectrum
  of \mbox{\decay{\Bp}{\Ds}{\Dzb}} decays in data, using a gradient boosted reweighter (GBR)~\cite{Rogozhnikov:2016bdp} technique.
The weights $w(\pt,y)$ are determined separately for Run~1 and Run~2.
Simulated \Bc events are corrected to match the measured linear dependence 
of $\inlfcfufd$ on \pt and $y$~\cite{LHCb-PAPER-2019-033}.
In addition, corrections using control samples are applied to the simulated events
to improve the agreement with data regarding particle identification (PID) variables,
the momentum scale and the momentum resolution.


\section{Candidate selection}
\label{sec:selection}
Charm-meson candidates are formed by combining two, three or four tracks that are incompatible with originating from any reconstructed PV.
The tracks are required to form a high-quality vertex and the scalar sum of their \pt must exceed $1.8\gevc$.
To reduce background from misidentified particles, the pion and kaon candidates must also satisfy loose criteria on \dllkpi, the ratio of the likelihood between the kaon and pion PID hypotheses.

The reconstructed mass of \Dz, \Ds and \Dp candidates is required to be within $\pm25\mevcc$ of their known values~\cite{PDG2020}.
For channels with a fully reconstructed \mbox{\decay{\Dpstar}{\Dz\pip}} meson, the mass difference $\Delta m$ between
the \Dpstar and the \Dz candidates is required 
to be within $\pm10\mevcc$ of the known value~\cite{PDG2020}.
If more than one charm-meson candidate is formed from the same track combination, 
only the best according to PID information is selected.

A \BporBc candidate is formed by combining a \DporDsorDstar candidate with a
\DzorDzb candidate if the combination has a \pt greater than $4.0\gevc$, forms a good-quality vertex
and originates from a PV. 
The reconstructed decay time of the charm meson candidates with respect to the \BporBc vertex divided by its uncertainty, $t/\sigma_t$,
 is required to exceed $-3$ for \Ds and \Dz mesons.
This requirement is increased to $+3$ for the longer-lived \Dp meson to eliminate background from \mbox{\decay{\Bp}{\Dzb\pip\pim\pip}} decays 
where the negatively charged pion is misidentified as a kaon.
Candidate \Bc decays that are compatible with the combination of
a fully reconstructed \decay{\B^0_{(s)}}{\DmorDsorDstar\pip(\pim\pip)} decay and a charged track are rejected.
To eliminate duplicate tracks, the opening angles between any pair of final-state particles are required to be at least 0.5\,\mrad.
The invariant-mass resolution of \BporBc decays is significantly improved by applying a kinematic fit~\cite{Hulsbergen:2005pu}
where the invariant masses of the \Dz and the \DporDsorDstar candidates are constrained to their known values~\cite{PDG2020},
all particles from the \DporDsorDstar, \Dz, and \BporBc decay are constrained to originate from their corresponding decay vertex
and the \BporBc candidate is constrained to originate from the PV
with which it has the smallest \chisqip.

To reduce the combinatorial background, while maintaining high efficiency for signal,
a multivariate selection based on a boosted decision tree (BDT)~\cite{Breiman,Roe} is employed.
The BDT classifier exploits kinematic and PID properties of selected candidates, namely:
the fit quality of the \BporBc candidate and both charm-meson candidate vertices;
the value of \chisqip of the \BporBc candidate;
the values of $t/\sigma_t$ of the \BporBc and both charm-meson candidates;
the reconstructed masses of the charm-meson candidates; and
the reconstructed masses of the pairs of opposite-charge tracks from the \DporDs candidate.
In addition, for each charm-meson candidate,
the smallest value of \pt and the smallest value of \chisqip among the decay products,
and the smallest (largest) value of \dllkpi among all kaon (pion) candidates,
are included as input variables for the BDT classifier.

The BDT training is performed separately
for the \Ds\DzorDzb, \Dp\DzorDzb  and \Dstarp\DzorDzb final states, separately for the \DKpi and \DKpipipi decay channels, and separately for the Run~1 and Run~2 data samples.
For a given \Dz final state, the same classifier is used for both \mbox{\decay{\Bc}{\DporDsorDstar\Dzb}}
and \mbox{\decay{\Bc}{\DporDsorDstar\Dz}} decays.
For signal decays, the BDT classifier is trained using simulated \Bc events,
while for background, data in the range $5350<m(\DporDsorDstar\DzorDzb)<6200\mevcc$ are used.
For the background sample,
the charm-meson mass windows are increased from $\pm25\mevcc$ to $\pm75\mevcc$,
to increase the size of the training sample.
The $k$-fold cross-training technique~\cite{geisser} with $k=5$ is used to avoid biases in the
calculation of the BDT output.

\begin{figure}[tb]
\includegraphics[width=7.8cm]{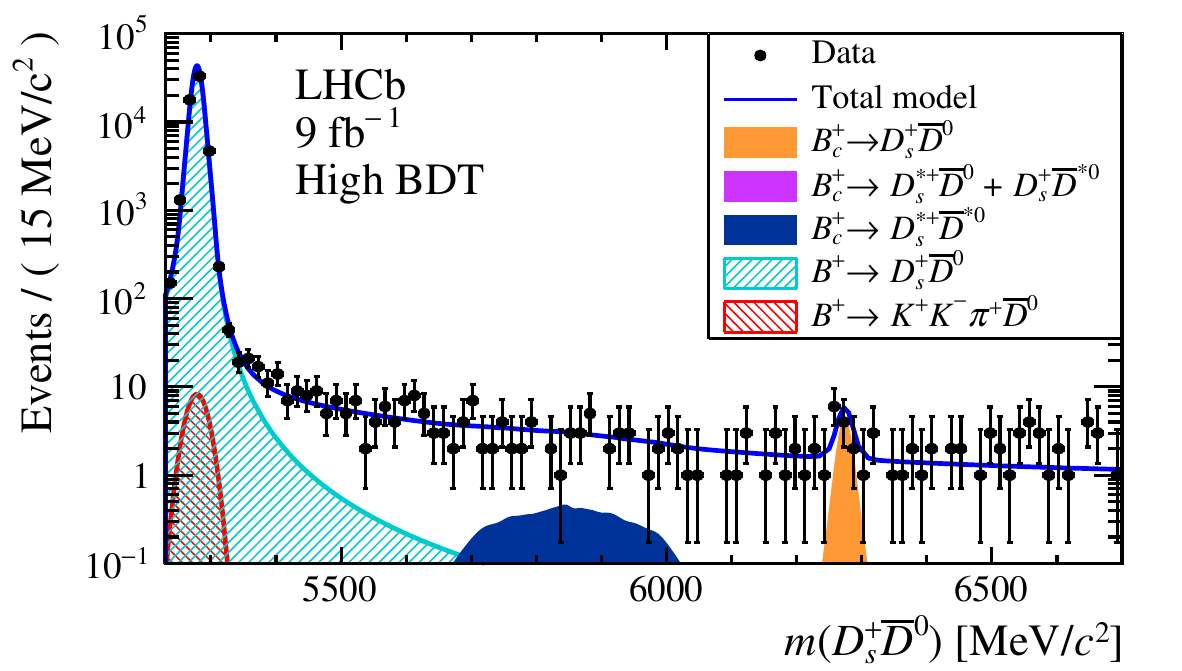}
\includegraphics[width=7.8cm]{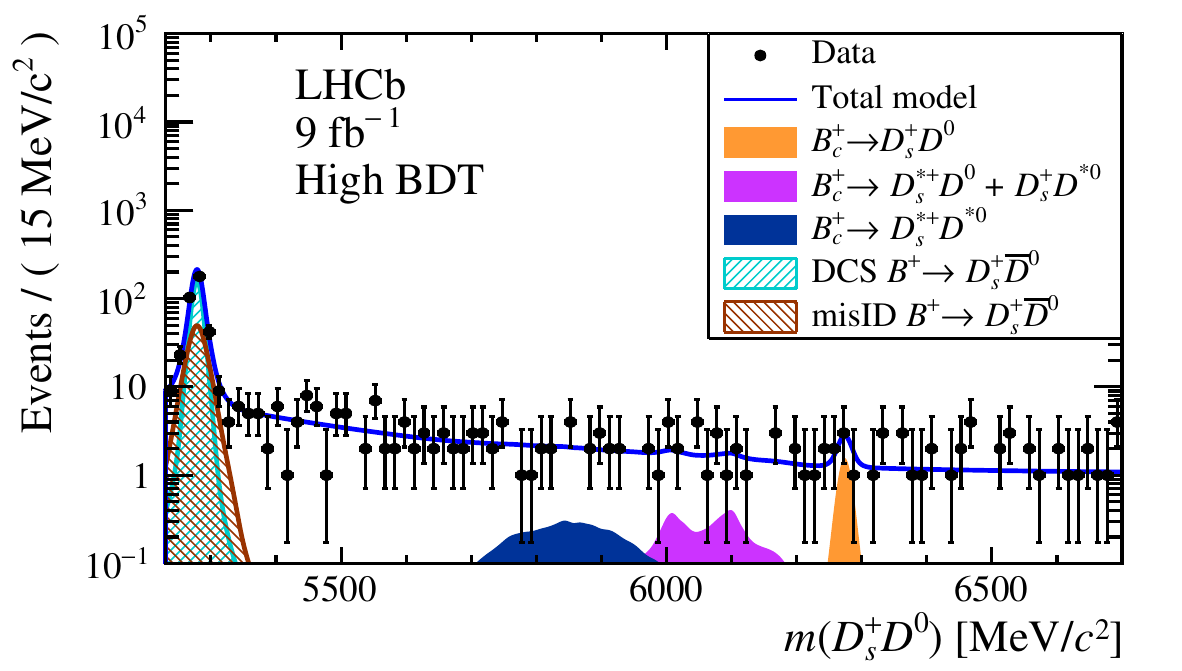}\\
\includegraphics[width=7.8cm]{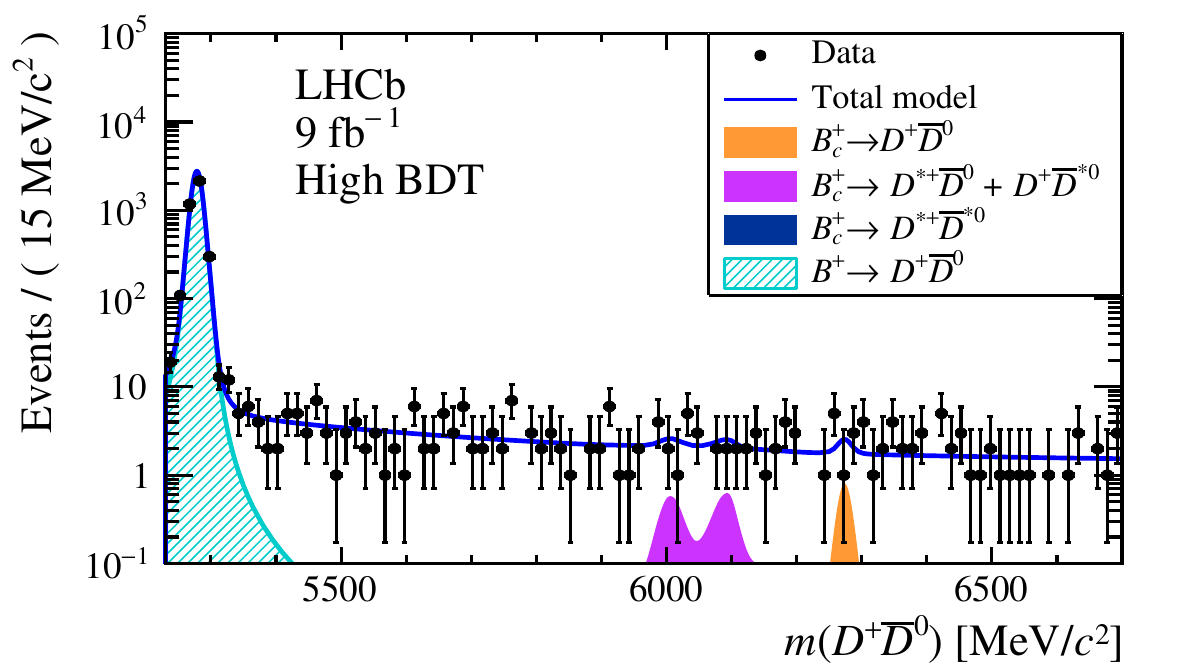}
\includegraphics[width=7.8cm]{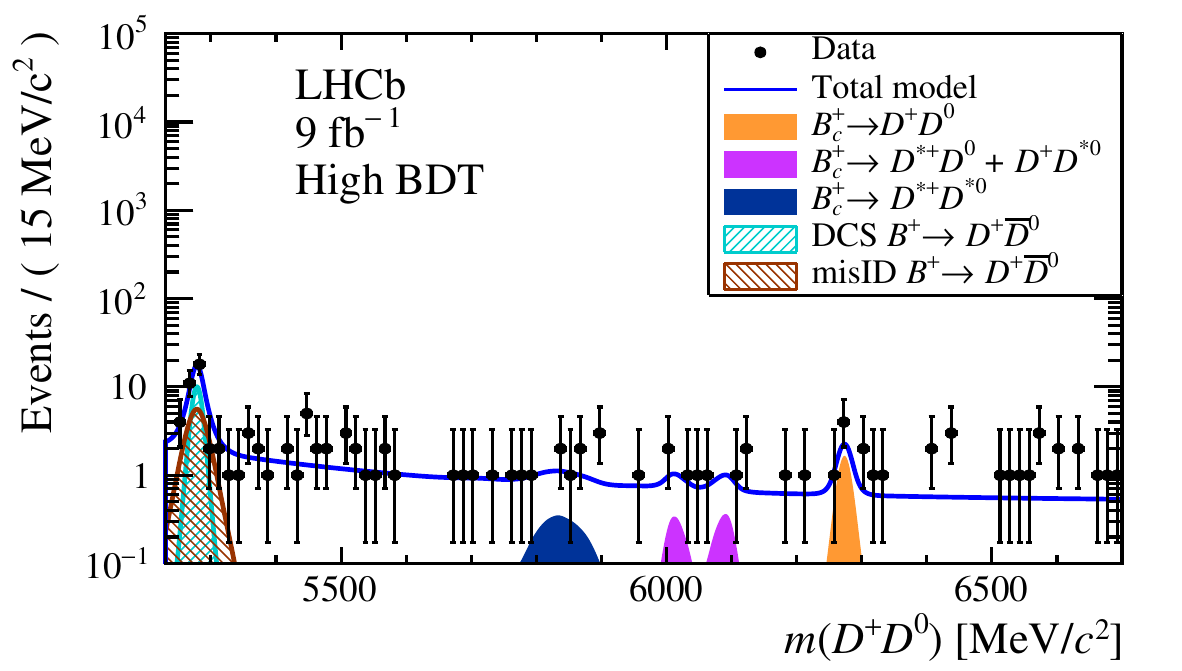}\\
\includegraphics[width=7.8cm]{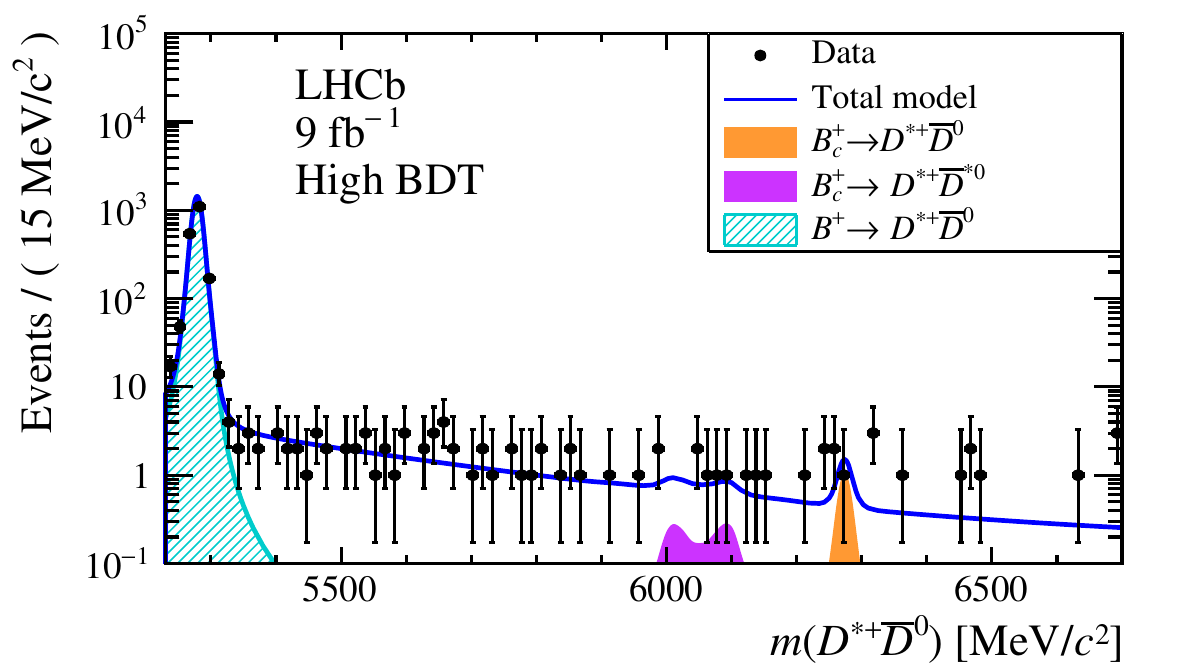}
\includegraphics[width=7.8cm]{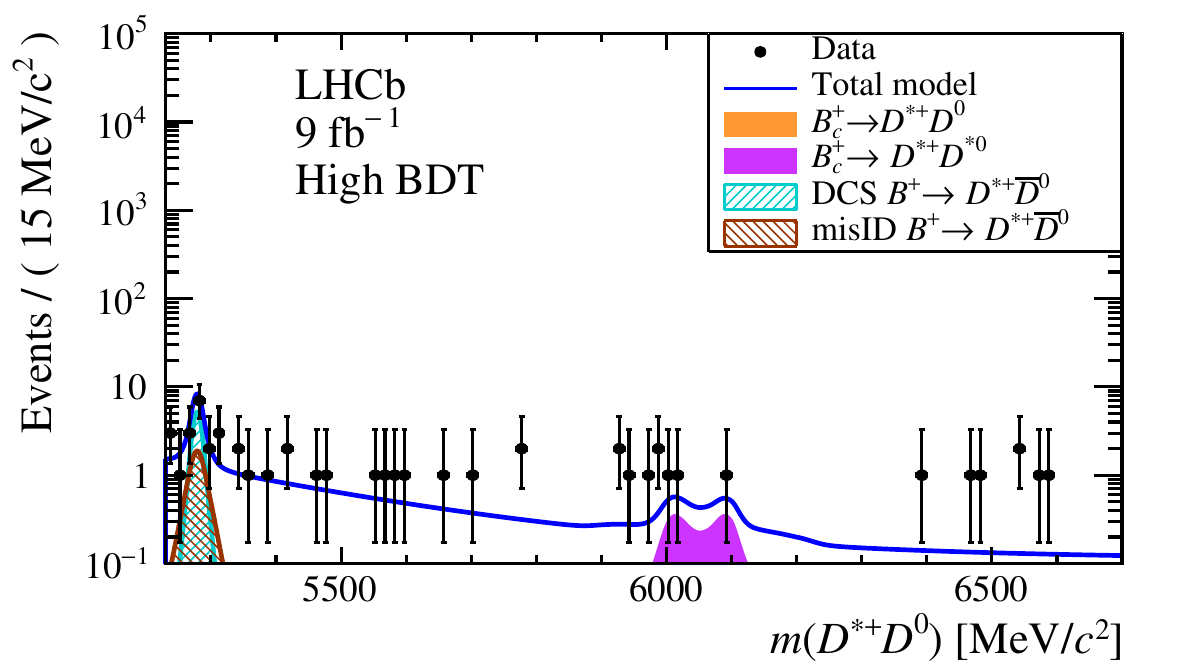}
\caption{Invariant-mass distributions for the selected \BporBc candidates in the highest BDT samples for (top left) \Ds\Dzb, (top right) \Ds\Dz, (center left) \Dp\Dzb, (center right) \Dp\Dz, (bottom left) \Dstarp\Dzb and (bottom right) \Dstarp\Dz, final states. The overlaid curves correspond to the sum of the corresponding fit results.}
\label{fig:massfits}
\end{figure}

The data are divided in increasing order of signal purity into three samples having low, medium and high BDT output.
Most of the sensitivity in this search comes from the data in the high BDT sample, but including data with lower signal purity increases the signal efficiency and constrains the shape of the combinatorial background.
A small fraction of the events ($\approx1\%$) have more than one \BporBc candidate that satisfies the minimum BDT requirement. 
In such cases, one randomly selected candidate is retained per event.
Figure~\ref{fig:massfits} shows the invariant mass distributions of selected \BporBc candidates in the highest BDT sample, summed over all \Dz final states.

\section{Model of the \BporBc mass distributions}
\label{sec:model}

To measure the signal yields, a model of the \BporBc candidate mass distribution is fitted to the data in the range $5230\leq m(\DporDsorDstar\DzorDzb)\leq 6700\mevcc$.
The model consists of the following components,
constrained to positive yields: 
the signals for fully reconstructed \Bp and \Bc decays;
the signal for \Bc decays with one missing \piz or photon;
the signal for \Bc decays with two missing \piz or photons;
the background from \mbox{\decay{\Bp}{\Dzb\Kp\Km\pip}} decays;
and the combinatorial background. 

Fully reconstructed \Bp and \Bc signals are described by the sum of a Gaussian function and
a Crystal Ball (CB)~\cite{Skwarnicki:1986xj} function,
extended to have power-law tails on both the low-mass
and the high-mass sides. 
The CB and Gaussian components share a common peak position.
The tail parameters of the CB and the ratio of the CB and Gaussian widths and integrals are determined from simulation, accounting for a dependence of both widths on the BDT output.
The ratio of the \Bc and \Bp widths is determined from simulation, 
while the overall width of the \Bp is a free parameter in the fit to data,
and is found to be consistent with the simulation.
The peak position of the \Bp signal is a free parameter in the fit to data,
and the known mass difference between the \Bp and the \Bc meson~\cite{PDG2020} is used to 
constrain the peak position of the \Bc signal.

Genuine \mbox{\decay{\Bp}{\DporDsorDstar\Dz}} decays are forbidden at tree level and consequently have a negligible yield, but doubly Cabibbo-suppressed (DCS) decays \mbox{\decay{\Dzb}{\Km\pip(\pim\pip)}}
result in crossfeed of \mbox{\decay{\Bp}{\DporDsorDstar\Dzb}} decays
in the \mbox{\DporDsorDstar\Dz} final state.
An additional source of crossfeed into the \mbox{\DporDsorDstar\Dz} final state is double misidentification of the pion and kaon in the Cabibbo-favoured \mbox{\decay{\Dz}{\Km\pip(\pim\pip)}} decay.
The DCS component is constrained in yield and shape by the large \mbox{\decay{\Bp}{\DporDsorDstar\Dzb}} signal, 
according to the known \Dz branching fractions~\cite{PDG2020}.
For the shape of the misidentified component, the width of the \mbox{\decay{\Bp}{\DporDsorDstar\Dzb}} peak
is scaled by a factor determined from a fit to \mbox{\decay{\Bp}{\Ds\Dz}} candidates, and also used for the \mbox{\decay{\Bp}{\DporDstar\Dz}} final states.
The yield of the misidentified component is a free parameter in all fits to data.

Models for decays with one or two missing neutral particles from \mbox{\DporDsorDstar} and/or \mbox{\DzorDzstar} decays are
implemented as templates, obtained from kernel fits~\cite{Cranmer:2001} to reconstructed mass distributions in simulation.
Both longitudinal and transverse polarisations of \mbox{\decay{\Bc}{\DporDsstar\DzorDzbstar}} decays contribute according to free polarisation fractions with different templates.

The Cabibbo-favoured \mbox{\decay{\Bp}{\Dzb\Kp\Km\pip}} decay is a background
to the \mbox{\decay{\Bp}{\Ds\Dzb}} channel, though its yield is strongly reduced by the \Ds mass requirement.
This background is modelled by a single Gaussian function, with the width determined from a sample of simulated decays and the 
normalisation determined from the peak in the \Bp mass distribution in the \Ds invariant mass sideband.

The combinatorial background is described by the sum of an exponential function and a constant,
where the parameters are allowed to differ between different \Dz 
decay modes, but are taken to be the same for all BDT samples of a 
given \Bc and \Dz decay channel.
Studies of the charm-meson invariant-mass sidebands
support these assumptions.

An unbinned extended maximum-likelihood fit is used to simultaneously describe 
the mass distributions of candidates in different BDT samples and 
different \Dz decay modes.
In these fits the background and \Bp yields vary independently,
but the branching fraction ratios $R$, $R'_{(+,0)}$ and $R''$,
defined in Eqs.~\ref{eq:bfcalc}--\ref{eq:bfcalc_2S},
are constrained to be identical between the BDT samples and \Dz decay modes.

\section{Systematic uncertainties}
\unboldmath
\label{sec:systematics}

Systematic uncertainties that can be expressed as a relative uncertainty on the branching fraction ratio are evaluated separately for Run 1 and Run 2, and for each \Bc decay, \Dz channel and BDT sample. 
Their effective contributions in the fit, calculated as a weighted average over BDT samples
and \Dz decay modes, are listed in Table~\ref{table:syst}.
    Where no uncertainty is given, this corresponds to either the absence of decays with two missing neutral particles in the \mbox{\Dstarp\DzorDzb} channel or the absence of the effect associated with an uncertainty in a given data-taking period or channel.

\begin{table}[tb]
    \caption{Effective contributions of the systematic uncertainties which are expressed as a relative uncertainty on the branching fraction ratio,
    combined over all BDT samples and \Dz decay modes, given in percent.}
    \begin{center}
\begin{tabular}{lSSSSSS}
\hline
Final state & \multicolumn{2}{c}{$\Ds\DzorDzb$} & \multicolumn{2}{c}{$\Dp\DzorDzb$} & \multicolumn{2}{c}{$\Dpstar\DzorDzb$}\\
                        & \multicolumn{1}{c}{Run 1} & \multicolumn{1}{c}{Run 2} & \multicolumn{1}{c}{Run 1} & \multicolumn{1}{c}{Run 2} & \multicolumn{1}{c}{Run 1} & \multicolumn{1}{c}{Run 2} \\
\hline
\Bc signal shape & 9.4 & 3.8 & 4.8 & 5.3 & 2.8 & 3.9 \\
\Bc production spectrum & 3.7 & 2.4 & 3.9 & 2.4 & 4.2 & 2.9 \\
\Bp production spectrum & 0.5 & 0.9 & 0.6 & 1.0 & 0.6 & 1.1 \\
Hit resolution parameterisation & \mbox{--} & 1.5 & \mbox{--} & 1.2 & \mbox{--} & 2.2 \\
$R$ simulation sample size & 1.2 & 1.0 & 1.4 & 1.1 & 1.5 & 1.5 \\
$R_{(+,0)}^\prime$ simulation sample size & 1.4 & 0.9 & 2.1 & 1.2 & 1.1 & 1.1 \\
$R^{\prime\prime}$ simulation sample size & 1.5 & 0.8 & 1.7 & 0.9 & \mbox{--} & \mbox{--} \\
\Bc lifetime & 1.3 & 1.4 & 1.3 & 1.3 & 2.1 & 2.6 \\
PID efficiencies & 1.6 & 1.2 & 2.8 & 0.8 & 2.2 & 1.4 \\
Multiple \BporBc candidates & 0.4 & 0.4 & 0.6 & 0.5 & 1.4 & 1.2 \\
Data-simulation differences & 0.1 & 0.1 & 0.1 & 0.1 & 0.1 & 0.2 \\
\decay{\Bp}{\Dzb\Kp\Km\pip} & 0.7 & 0.5 & \mbox{--} & \mbox{--} & \mbox{--} & \mbox{--} \\
$\BF(\decay{\Dstarp}{\Dp\pizgamma})$ & \mbox{--} & \mbox{--} & 1.5 & 1.5 & \mbox{--} & \mbox{--} \\
\hline
$R$ total & 10.4 & 5.3 & 7.2 & 6.6 & 6.3 & 6.5 \\
$R_{(+,0)}^\prime$ total & 4.6 & 3.7 & 5.7 & 3.8 & 5.5 & 5.0 \\
$R^{\prime\prime}$ total & 4.6 & 3.7 & 5.5 & 3.7 & \mbox{--} & \mbox{--} \\
\hline
\end{tabular}
\end{center}
\label{table:syst}
\end{table}
The uncertainty on the \Bc signal shape is evaluated by changing the \Bc signal shape to the sum of two Gaussian functions,
and evaluating the median fractional change of the measured yield in pseudoexperiments performed with a background-only model.
Uncertainties related to the $\Bc$ production spectrum are evaluated by changing the slope parameters from Ref.~\cite{LHCb-PAPER-2019-033} by their quoted uncertainties.
The uncertainty on the \pt- and $y$-dependent weights used to correct the \Bp production spectrum in simulation is estimated by changing the settings of the GBR algorithm.
Hit resolution parameterisation in the silicon vertex detector affects the \chisqip distribution.
The uncertainty associated with the parameterisation is therefore evaluated with simulation by varying the minimal value of the \chisqip applied to the final-state tracks.

The limited size of the simulation samples results in uncertainties that are uncorrelated between the BDT samples and \Dz decay channels on the efficiency ratios $\varepsilon(\Bc)/\varepsilon(\Bp)$. All other systematic uncertainties are treated as fully correlated.
A small uncertainty on the reconstruction efficiency results from that on the \Bc lifetime~\cite{PDG2020}.
Uncertainties on the PID efficiencies cancel to first order in the ratio $\varepsilon(\Bc)/\varepsilon(\Bp)$ because of the identical particle content of the final state, and the difference in relative efficiencies with and without PID corrections is used to estimate the uncertainty from the PID correction procedure.
 The requirement to select at most one \BporBc candidate per event introduces an efficiency that may not be well reproduced by simulation. Therefore, the fraction of candidates removed by the requirement of at most one \BporBc candidate per event is attributed as a systematic uncertainty.
 Residual differences appear in the comparison of the distributions of the BDT output between background-subtracted \Bp signal from data and simulation. The effect on the relative efficiency is evaluated by correcting the simulation to match the distributions in data.
The background from \mbox{\decay{\Bp}{\Dzb\Kp\Km\pip}} decays to the  \mbox{\decay{\Bp}{\Ds\Dzb}} signal is assigned an uncertainty of 100\% of its yield, resulting in a fractional uncertainty of less than 1\%.
The measurements of the branching fraction ratios according to Eqs.~\ref{eq:bfcalc_1SPDpDs} and \ref{eq:bfcalc_2S} involve the value of \mbox{$\BF(\decay{\Dstarp}{\Dp\pizgamma})$}, the uncertainty of which~\cite{PDG2020} is taken into account.

    Other uncertainties, listed below, are instead taken into account by varying the fit model.
    Unless specified otherwise, these uncertainties are taken into account by replacing fixed values of the model parameters by their Gaussian constraints.

The uncertainty on the combinatorial background shape is evaluated by considering 
    a single exponential function as an alternative to the exponential plus constant model, implemented using the discrete profiling method~\cite{Dauncey:2015}.
    The \Bp shape uncertainty has a negligible effect on the \Bp yield but, because of its long tails, 
    results in an uncertainty on the background shape. The effect is evaluated by assigning an uncertainty
    on the tail parameters determined from a fit to simulated events.
    The uncertainty on the $\alpha$ parameters of the CB function
    is increased by adding in quadrature the largest observed difference between data and simulation of this parameter in \mbox{\decay{\Bp}{\Ds\Dzb}} decays.
    The fractional uncertainty on the yield of DCS crossfeed is the sum in quadrature of the fractional uncertainty on the normalisation yield and on the branching fractions \mbox{$\BF(\decay{\Dz}{\Kp\pim(\pip\pim))}$}~\cite{PDG2020}.
    The uncertainty on the difference between the \Bc and \Bp peak positions, $0.5\mevcc$,
    arises due to uncertainty on the measured masses~\cite{PDG2020,LHCb-PAPER-2020-003}
and the momentum scale uncertainty~\cite{LHCb-PAPER-2012-048}.
The uncertainty on the ratio of the \Bc to \Bp invariant-mass resolution is determined from the statistical uncertainties of the fits to simulated decays.
    Statistical uncertainties in the templates of \mbox{\decay{\Bc}{\DporDsstar\DzorDzb}} and \mbox{\decay{\Bc}{\DporDs\DzorDzbstar}} decays with one missing neutral pion or photon are accounted for by allowing a small contribution from the other template.
        The \mbox{\decay{\Bc}{\DporDsstar\DzorDzbstar}} signals have contributions from 
    both transverse and longitudinal polarisations,
    which have differently shaped distributions of the reconstructed mass.
    This is accounted for by evaluating upper limits both for fully longitudinal and fully transverse polarisations, and reporting the least stringent upper limit.
    Including all model uncertainties results in an increase of the
    upper limits on the branching fraction ratios, discussed in Sec.~\ref{sec:results}, of 7\% on average.

\section{Results and conclusions}
\unboldmath
\label{sec:results}

\begin{figure}[tb]
\includegraphics[width=7.8cm]{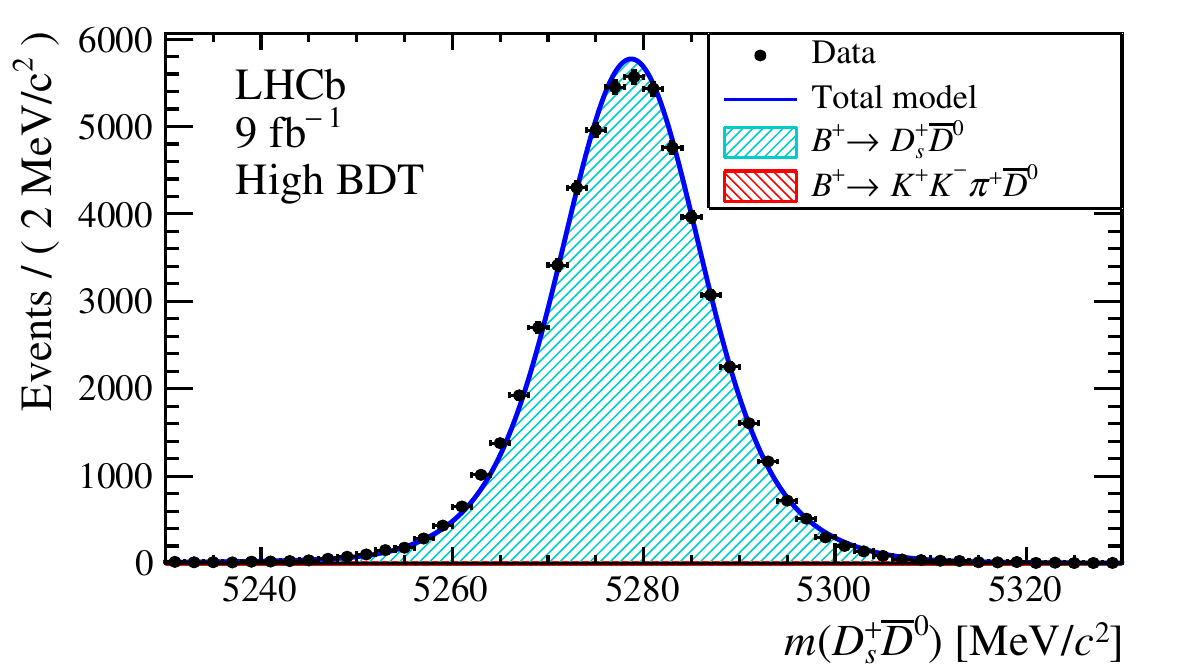}
\includegraphics[width=7.8cm]{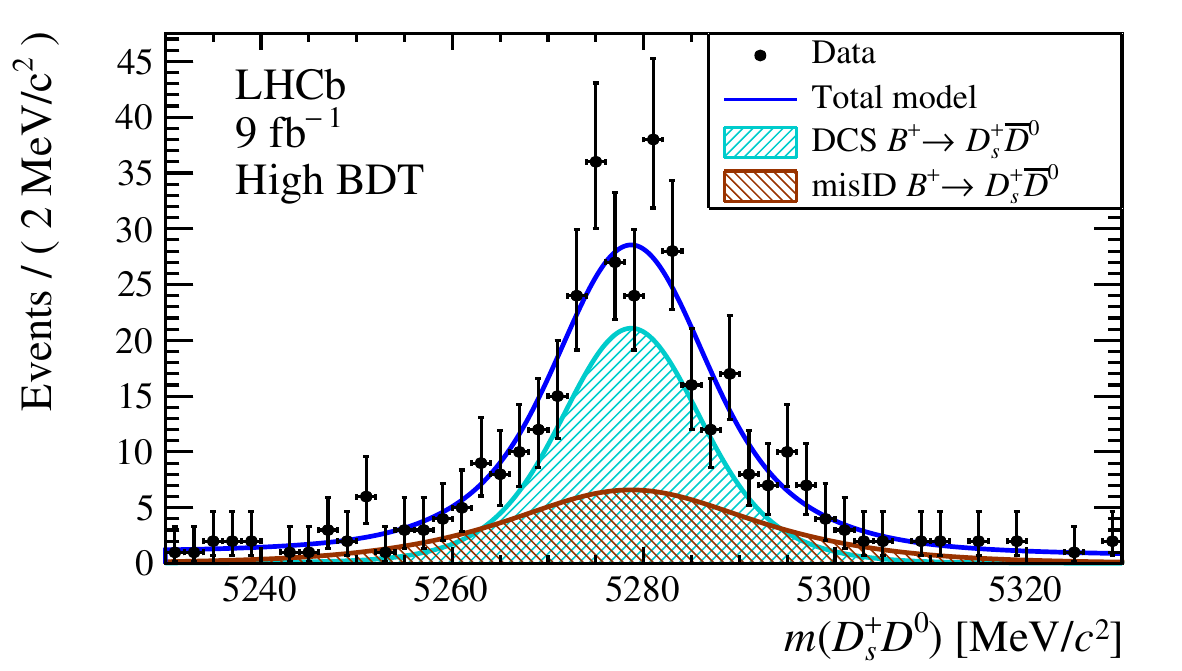}
\caption{Invariant-mass distributions for the selected \Bp candidates in the highest BDT samples, in the region near the \Bp mass, for (left) \Ds\Dzb and (right) \Ds\Dz final states.
The overlaid curves correspond to the sum of the corresponding fit results.}
\label{fig:massfits_Bp}
\end{figure}

To determine the \Bc branching fraction ratios $R$, $R'_{(+,0)}$ and $R''$, 
fits to data are performed separately for the six \Bc final states
and for Run~1 and Run~2, while different \Dz decay modes and BDT samples are fit simultaneously.
The results of the fits are shown in Fig.~\ref{fig:massfits},
where the data of the highest BDT samples and the corresponding fit results are summed over the \Dz decay channels and over data-taking periods.
Detailed views of the \Ds\DzorDzb final states near the \Bp mass
are shown in Fig.~\ref{fig:massfits_Bp} 
which validate the model of the large
\mbox{\decay{\Bp}{\Ds\Dzb}} signal and its crossfeed to the \Ds\Dz final state. 
The integrals of the fit results in a $\pm40\mevcc$ window around the \Bp mass differ less then 0.2\% from the candidate counts.

The significance of the \Bc signals are calculated using Wilks' theorem~\cite{Wilks:1938dza} as
\mbox{$S=\sqrt{2\Delta\log\mathcal{L}}$}, where $\Delta\log\mathcal{L}$ is  the difference in the logarithm of the likelihood between the signal plus background and  background-only hypotheses.
Systematic uncertainties are included in the calculation of the significance
through nuisance parameters in a minimised profile likelihood.

Evidence is found only for the decay \mbox{\decay{\Bc}{\Ds\Dzb}} in Run~2 data, with a significance of 3.7 standard deviations, and the measured branching fraction ratio is
\mbox{$R(\Ds\Dzb)=(3.6^{+1.5+0.3}_{-1.2-0.2})\times 10^{-4}$}, where the first uncertainty is statistical and the second is systematic.
The quoted significance for this channel is compatible with estimates from simulated pseudoexperiments.

The values of $R$, $R^\prime_{(+,0)}$ and $R^{\prime\prime}$ from Run~1 and Run~2 cannot be directly combined since the value of \inlfcfu depends on the $pp$ centre-of-mass energy.
Therefore, a combined fit of both the Run~1 and Run~2 data sets is made to the absolute \Bc branching fractions, 
using external input for $\BF(\decay{\Bp}{\Ds\Dzb})$, 
$\BF(\decay{\Bp}{\Dp\Dzb})$,
$\BF(\decay{\Bp}{\Dstarp\Dzb})$~\cite{PDG2020},
and \inlfcfu~\cite{LHCb-PAPER-2019-033}, which depends on the theory prediction of \mbox{$\BF(\decay{\Bc}{\jpsi\mup\neum})$}.
Corresponding uncertainties are included.
In this combined fit the excess for \mbox{\decay{\Bc}{\Ds\Dzb}} 
has a significance of 3.4 standard deviations,
or 2.5 standard deviations when considering the probability
of the excess to appear in any of the sixteen final states considered.
The corresponding value of the branching fraction is 
\mbox{$\BF(\decay{\Bc}{\Ds\Dzb})=(3.5^{+1.5+0.3}_{-1.3-0.2}\pm1.0)\times 10^{-4}$},
where the first uncertainty is statistical, the second systematic and the third due to external input.

Upper limits are reported on
the ratio of branching fractions for all decays,
calculated at 90\% and 95\% confidence level (C.L.) with the frequentist CL$_s$ method~\cite{Read:2002,GammaCombo},
separately for Run~1 and Run~2.
These limits are listed in Table~\ref{table:limits}.
Limits for Run~1 are tighter than in Ref.~\cite{LHCb-PAPER-2017-045} in particular for $R'_{(+,0)}$ and $R''$, mainly because of better constraints on the shape of the combinatorial background.
\begin{table}[tb]
    \caption{Upper limits on the branching fraction ratios $R$, $R_{(+,0)}'$ and $R''$ of \Bc to \Bp decays,
    defined in Eqs.~\ref{eq:bfcalc}--\ref{eq:bfcalc_2S}, at the 90(95)\% C.L.  for Run~2 and Run~1 data, 
    in units of $10^{-3}$.}
    \begin{center}\begin{tabular}{lcc}
        \hline
        & Run~2 & Run~1 \\
        & $6\invfb$, $13\tev$ & $3\invfb$, 7~and $8\tev$\\
        \hline
      \vspace{-0.45cm} \\
$R(\Ds\Dzb)$& $0.57\,(0.62)$ & $0.45\,(0.58)$ \\
$R_+'(\Ds\Dzb)$& $0.36\,(0.42)$ & $0.45\,(0.75)$ \\
$R_0'(\Ds\Dzb)$& $0.27\,(0.36)$ & $0.64\,(0.71)$ \\
$R''(\Ds\Dzb)$& $0.9\,(1.1)$ & $1.1\,(1.5)$ \\
$R(\Ds\Dz)$& $0.22\,(0.25)$ & $0.51\,(0.62)$ \\
$R_+'(\Ds\Dz)$& $0.59\,(0.72)$ & $0.76\,(0.89)$ \\
$R_0'(\Ds\Dz)$& $0.49\,(0.60)$ & $0.74\,(0.88)$ \\
$R''(\Ds\Dz)$& $0.9\,(1.0)$ & $1.6\,(2.3)$ \\
$R(\Dp\Dzb)$& $3.5\,(4.4)$ & \hspace{1ex}$8\,(11)$ \\
$R_+'(\Dp\Dzb)$& $26\,(33)$ & $45\,(52)$ \\
$R_0'(\Dp\Dzb)$& $11\,(12)$ & $16\,(20)$ \\
$R''(\Dp\Dzb)$& $21\,(28)$ & $90\,(110)$ \\
$R(\Dp\Dz)$& $2.9\,(3.6)$ & $12\,(14)$ \\
$R_+'(\Dp\Dz)$& $18\,(21)$ & $17\,(33)$ \\
$R_0'(\Dp\Dz)$& $6.6\,(7.3)$ & $8.2\,(9.8)$ \\
$R''(\Dp\Dz)$& $17\,(20)$ & $82\,(94)$ \\
$R(\Dstarp\Dzb)$& $6.9\,(8.4)$ & $26\,(28)$ \\
$R'(\Dstarp\Dzb)$& $16\,(19)$ & $31\,(44)$ \\
$R(\Dstarp\Dz)$& $3.6\,(4.4)$ & \hspace{1ex}$8\,(13)$ \\
$R'(\Dstarp\Dz)$& $12\,(15)$ & $36\,(45)$ \\
            \hline
        \end{tabular}
    \end{center}
    \label{table:limits}
\end{table}

Upper limits on the absolute \Bc branching fractions are based on the Run 2 dataset alone, which has nearly four times the sensitivity of the Run 1 dataset.
The upper limits at 90(95\%) C.L. are
\begin{align*}
    \BF(\decay{\Bc}{\Ds\Dzb}) < 7.2\,(8.4)\times 10^{-4};\\
    \BF(\decay{\Bc}{\Ds\Dz}) < 3.0\,(3.7)\times 10^{-4};\\
    \BF(\decay{\Bc}{\Dp\Dzb}) < 1.9\,(2.5)\times 10^{-4};\\
    \BF(\decay{\Bc}{\Dp\Dz}) < 1.4\,(1.8)\times 10^{-4};\\
    \BF(\decay{\Bc}{\Dss\Dzb}) < 5.3\,(5.7)\times 10^{-4};\\
    \BF(\decay{\Bc}{\Ds\Dstarzb}) < 4.6\,(5.6)\times 10^{-4};\\
    \BF(\decay{\Bc}{\Dss\Dz}) < 0.9\,(1.0)\times 10^{-3};\\
    \BF(\decay{\Bc}{\Ds\Dstarz}) < 6.6\,(8.4)\times 10^{-4};\\
    \BF(\decay{\Bc}{\Dstarp\Dzb}) < 3.8\,(4.8)\times 10^{-4};\\
    \BF(\decay{\Bc}{\Dstarp\Dz}) < 2.0\,(2.4)\times 10^{-4};\\
    \BF(\decay{\Bc}{\Dp\Dstarzb}) < 6.5\,(8.2)\times 10^{-4};\\
    \BF(\decay{\Bc}{\Dp\Dstarz}) < 3.7\,(4.6)\times 10^{-4};\\
    \BF(\decay{\Bc}{\Dss\Dstarzb}) < 1.3\,(1.5)\times 10^{-3};\\
    \BF(\decay{\Bc}{\Dss\Dstarz}) < 1.3\,(1.6)\times 10^{-3};\\
    \BF(\decay{\Bc}{\Dstarp\Dstarzb}) < 1.0\,(1.3)\times 10^{-3};\\
    \BF(\decay{\Bc}{\Dstarp\Dstarz}) < 7.7\,(8.9)\times 10^{-4}.\\
\end{align*}
The reported upper limits on \Bc decays with a \Dstarp meson in the final state
are based on the analyses of fully reconstructed \decay{\Dstarp}{\Dz\pip} decays, which have a higher sensitivity
than the channels with partially reconstructed \decay{\Dstarp}{\Dp\pizgamma} decays.

In conclusion, this article reports the results of a search 
for \mbox{\decay{\Bc}{\DporDsorDstar}{\DzorDzb}} decays,
covering sixteen \Bc decay channels,
which include partially reconstructed decays where one or two neutral pions or photons from the decay of an excited charm meson are not reconstructed.
The results, based on $pp$ collision data corresponding to $9\invfb$ of integrated luminosity,
supersede an earlier LHCb measurement~\cite{LHCb-PAPER-2017-045}
on Run~1 data only.
No signal is observed in any of the channels investigated,
consistent with the Standard Model expectation. 
An excess with a significance of 3.4 standard deviations is found
for the decay \mbox{\decay{\Bc}{\Ds\Dzb}},
 which is in tension with the theoretical expectation~\cite{Rui:2012qq,Kiselev:2003ds,Ivanov:2002un,Ivanov:2006ni}.

%% file: acknowledgements.tex
\section*{Acknowledgements}
%
%
\noindent We express our gratitude to our colleagues in the CERN
accelerator departments for the excellent performance of the LHC. We
thank the technical and administrative staff at the LHCb
institutes.
We acknowledge support from CERN and from the national agencies:
CAPES, CNPq, FAPERJ and FINEP (Brazil); 
MOST and NSFC (China); 
CNRS/IN2P3 (France); 
BMBF, DFG and MPG (Germany); 
INFN (Italy); 
NWO (Netherlands); 
MNiSW and NCN (Poland); 
MEN/IFA (Romania); 
MSHE (Russia); 
MICINN (Spain); 
SNSF and SER (Switzerland); 
NASU (Ukraine); 
STFC (United Kingdom); 
DOE NP and NSF (USA).
We acknowledge the computing resources that are provided by CERN, IN2P3
(France), KIT and DESY (Germany), INFN (Italy), SURF (Netherlands),
PIC (Spain), GridPP (United Kingdom), RRCKI and Yandex
LLC (Russia), CSCS (Switzerland), IFIN-HH (Romania), CBPF (Brazil),
PL-GRID (Poland) and NERSC (USA).
We are indebted to the communities behind the multiple open-source
software packages on which we depend.
Individual groups or members have received support from
ARC and ARDC (Australia);
AvH Foundation (Germany);
EPLANET, Marie Sk\l{}odowska-Curie Actions and ERC (European Union);
A*MIDEX, ANR, IPhU and Labex P2IO, and R\'{e}gion Auvergne-Rh\^{o}ne-Alpes (France);
Key Research Program of Frontier Sciences of CAS, CAS PIFI, CAS CCEPP, 
Fundamental Research Funds for the Central Universities, 
and Sci. \& Tech. Program of Guangzhou (China);
RFBR, RSF and Yandex LLC (Russia);
GVA, XuntaGal and GENCAT (Spain);
the Leverhulme Trust, the Royal Society
 and UKRI (United Kingdom).

%% file: Authorship_LHCb-PAPER-2021-023.tex
\centerline
{\large\bf LHCb collaboration}
\begin
{flushleft}
\small
R.~Aaij$^{32}$,
A.S.W.~Abdelmotteleb$^{56}$,
C.~Abell{\'a}n~Beteta$^{50}$,
T.~Ackernley$^{60}$,
B.~Adeva$^{46}$,
M.~Adinolfi$^{54}$,
H.~Afsharnia$^{9}$,
C.~Agapopoulou$^{13}$,
C.A.~Aidala$^{86}$,
S.~Aiola$^{25}$
% ...

{\footnotesize \it

% ...
$^{9}$Universit{\'e} Clermont Auvergne, CNRS/IN2P3, LPC, Clermont-Ferrand, France\\
% ...
$^{13}$LPNHE, Sorbonne Universit{\'e}, Paris Diderot Sorbonne Paris Cit{\'e}, CNRS/IN2P3, Paris, France\\
% ...
$^{25}$INFN Sezione di Milano, Milano, Italy\\
% ...
$^{32}$Nikhef National Institute for Subatomic Physics, Amsterdam, Netherlands\\
% ...
$^{46}$Instituto Galego de F{\'\i}sica de Altas Enerx{\'\i}as (IGFAE), Universidade de Santiago de Compostela, Santiago de Compostela, Spain\\
% ...
$^{50}$Physik-Institut, Universit{\"a}t Z{\"u}rich, Z{\"u}rich, Switzerland\\
% ...
$^{54}$H.H. Wills Physics Laboratory, University of Bristol, Bristol, United Kingdom\\
% ...
$^{56}$Department of Physics, University of Warwick, Coventry, United Kingdom\\
% ...
$^{60}$Oliver Lodge Laboratory, University of Liverpool, Liverpool, United Kingdom\\
% ...
$^{68}$Syracuse University, Syracuse, NY, United States\\
% ...
$^{86}$University of Michigan, Ann Arbor, United States, associated to $^{68}$\\
\bigskip
% ...
\medskip
}
\end{flushleft}